\documentclass[twocolumn,aps,amsmath,amssymb,prb]{revtex4}

\usepackage{graphicx}
\usepackage{latexsym}
\usepackage{amssymb}

\newcommand{\BE}{\begin{equation}}
\newcommand{\EE}{\end{equation}}
\newcommand{\BA}{\begin{eqnarray}}
\newcommand{\EA}{\end{eqnarray}}
\usepackage{graphicx}
\usepackage{dcolumn}
\usepackage{bm}
\input epsf.tex

\def\sn{{\smallskip\par\noindent}}

\begin{document}

\begin{flushleft}
\begin{Large}
{\bf\sf{Flat lens imaging does not need negative refraction}}
\end{Large}

\sn

{\bf\small\sf{Chao-Hsien Kuo and Zhen Ye}} \\
 {\scriptsize\sl Wave Phenomena Laboratory, Department of
Physics, National Central University, Chungli, Taiwan 320,
Republic of China.\\ (cond-mat/0312288; an extended version will
appear in J. Appl. Phys.)}
\end{flushleft}

\begin{small}
Sir, in a recent communication Parimi et al. \cite{Nature}
reported the experimental results on imaging by a flat lens made
of photonic crystals. They attributed the observed focusing to
all-angle negative refraction, which may be expected for the
Left-Handed-Materials (LHMs). Here we demonstrate that the
experimental observation is {\it irrelevant} to all-angle negative
refraction. Rather, the phenomenon is a natural result of the
anisotropic scattering by an array of scatterers.

We consider the system in the experiment\cite{Nature}: the
two-dimensional flat slab made of a photonic crystal fabricated
from a square array of cylindrical alumina rods. All the
parameters are taken exactly from the experiment\cite{Nature}. The
dielectric constant of the rods is 9.2 and the lattice constant is
1.8 cm\cite{Nature}. Two arrangements of the cylinders are
considered. The first is exactly identical to that in the
experiment. A slab has the size of 10x19 lattice constant. The
microwave source is placed at a distance of 2.25 cm from the left
side of the slab. The frequency is taken as 9.3 GHz. As the
comparison, we also consider another arrangement: the square array
is added with one more layer of cylinders to become 10x20, and the
source is moved upward by a half lattice constant. The transmitted
wave from the source is scattered multiply by the cylinders and
the scattering can be solved {\it exactly} by the standard
multiple scattering theory\cite{MST}.

\begin{figure}[hbt]
\begin{center}
\epsfxsize=2.5in \epsffile{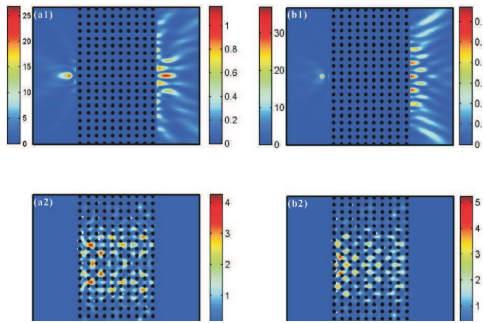} \caption{\label{fig1}\small
\sf The image of the intensity-fields for flat slabs. (a1) and
(b1) refer to two arrangements. In (a1), all the arrangements
including the source, the field resolution, and the cylinders are
identical to the experiment\cite{Nature}. (a2) and (b2) plot the
fields inside the slabs corresponding to those in (a1) and (b1)
respectively. The scales for the fields for the left and right
sides are shown in the figure.}
\end{center}
\end{figure}

Figure~\ref{fig1} shows the imaging fields across the slabs. In
(a1) and (b1), we plot the fields outside the slabs, and the
fields inside the corresponding slabs are plotted in (a2) and
(b2), with (a) referring to the case in the experiment. The result
in (a1) remarkably reproduces the experimental observation. A
focused image is obviously seen on the far side. Parimi et
al.\cite{Nature} suggested ``... to focus a diverging beam from a
point source, the material must exhibit all-angle negative
refraction, ..." We find that this perception is invalid. If the
image were caused by the all angle-negative refraction, one would
expect that (1) another focused image should exist inside the
slab; (2) both images should not be sensitive to the lattice size,
nor the source location. Our simulation shows that there is no
image inside the slab and the image on the far side is sensitive
to both the size of photonic crystal slab and the location of the
source. This has been clearly depicted by Fig.~\ref{fig1}(a2),
(b1) and (b2). Fig.~\ref{fig1}(a2) shows the image field inside
the slab used in the experiment. No focused image prevails.
Rather, the waves more or less go straight across the slab. For
the second slab, the image field on the far side is changed
completely, while the general features of the field inside the
slab remain unchanged.

The focusing effect by the flat slab in Fig.~\ref{fig1}(a1) is
caused by the anisotropic scattering of the regular arrays of
cylinders. Figure~\ref{fig2} shows the band structure result and
the transmission across the slab used in the
experiment\cite{Nature}. The simulation setup complies with the
experimental arrangement listed on the webpage\cite{NEU}. The
intensity transmission in the two principal directions, i.~e.
$\Gamma X$ and $\Gamma M$, is plotted. The result for the $\Gamma
X$ direction agrees with the experimental result\cite{NEU}. We
note that the dielectric constant used in the transmission
experiment\cite{NEU} is 8.9, while it is 9.2 in the imaging
experiment\cite{Nature}. To be consistent with the imaging
experiment, we use 9.2 throughout. It is seen that the
transmission in the $\Gamma X$ direction is much stronger than
that in the $\Gamma M$ direction at the experimental frequency 9.3
GHz. Due to the anisotropic scattering, the waves rather prefer to
travel along the $\Gamma X$ than along the $\Gamma M$ direction.
Such an anisotropy-caused imaging phenomenon has also been
discussed for frequencies located in the partial bandgap regimes
in a variety of situations\cite{neg2}.

\begin{figure}[hbt]
\begin{center}
\epsfxsize=2in \epsffile{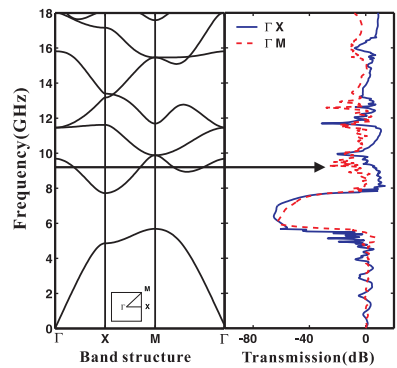} \caption{ \label{fig2}\small\sf
Left panel: The band structure calculation. Right panel: the
transmission calculation; the solid line refers to the result for
the transmission along the  $\Gamma X$ direction and the dotted
line to that from the $\Gamma M$ direction. }
\end{center}
\end{figure}

\end{small}

\vskip 12pt
\renewcommand{\refname}{}
\makeatletter
\renewcommand{\@biblabel}[1]{\hfill#1.}
\makeatother \vspace{-1.2cm}

\begin{scriptsize}

{Correspondence and requests for materials should be addressed to
Z.Y. (e-mail: zhen@shaw.ca).}

\end{scriptsize}

\end{document}